# Can Synergy in Triple-Helix Relations be Quantified?
# A Review of the Development of the Triple-Helix Indicator

Loet Leydesdorff[a] & Han Woo Park[b]

**Abstract**

Triple-Helix arrangements of bi- and trilateral relations can be considered as adaptive eco-systems. During the last decade, we have further developed a Triple-Helix indicator of synergy as reduction of uncertainty in niches that can be shaped among three or more distributions. Reduction of uncertainty can be generated in *correlations* among distributions of relations, but this (next-order) effect can be counterbalanced by uncertainty generated in the *relations*. We first explain the indicator, and then review possible results when this indicator is applied to (i) co-author networks of academic, industrial, and governmental authors and (ii) synergies in the distributions of firms over geographical addresses, technological classes, and industrial-size classes for a number of nations. Co-variation is then considered as a measure of relationship. The balance between globalizing and localizing dynamics can be quantified. Too much synergy locally can also be considered as lock-in. Tendencies are different for the globalizing knowledge dynamics versus locally retaining wealth from knowledge in industrial innovations.

[a] University of Amsterdam, Amsterdam School of Communication Research (ASCoR), Kloveniersburgwal 48, 1012 CX Amsterdam, The Netherlands; loet@leydesdorff.net
[b] Department of Media & Communication, YeungNam University, 214-1, Dae-dong, Gyeongsan-si, Gyeongsangbuk-do, South Korea, 712-749; hanpark@ynu.ac.kr.

# 1. Introduction

Whereas a political economy is based on two organizing principles—the economy and politics—a knowledge-based economy is more complex to the extent that technological developments can be expected continuously to upset the equilibrium of the market. The third dynamic of organized novelty production has to be endogenized into a model of technological innovations (Nelson & Winter, 1982; Storper, 1997). Schumpeter (1936, at p. 62) first defined innovation in relation to the two dynamics of factor substitutions along the production function and shifts of the production function towards the origin. Two dynamics can mutually shape each other and thus form a trajectory; a trajectory can lead to a competitive edge (Leydesdorff & Van den Besselaar, 1998). Three dynamics, however, shape an adaptive eco-system because the additional—third—degree of freedom allows for both globalization and localization. A regime at the global level and potentially synergetic niches in local optima can then be expected (Dolfsma & Leydesdorff, 2009). The systems dynamics is adaptive to relations that are added or deleted.

Under what conditions can the Triple Helix (TH) of university-industry-government (UIG) relations be considered as an eco-system with synergy? Etzkowitz & Leydesdorff (1995; 2000) considered the synergy in a Triple Helix system as the result of an overlay of exchanges among perspectives on the bi- and trilateral relations. This overlay is constructed in terms of relations, but can be expected to add another (hyper-cyclic) dynamic of shared meanings to the events (Leydesdorff, 1994, pp. 186f.). Thus, one obtains in the case of three or more distributions a double-layered system: at the bottom a network of exchange relations, and a next layer in which these relations are appreciated from different perspectives. Next-order sharing of perspectives—that is, translations among codes of communication—can feedback on or feed-forward to the relations in the network.

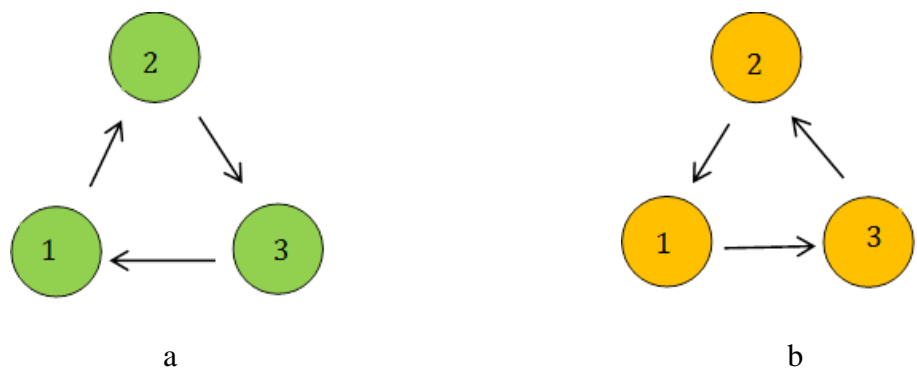

**Figure 1:** Circulation and feedback in cycles in two directions.

As against a double helix, symmetry can be broken in a three- (or more-)dimensional system and the dynamics can therefore be positive or negative (Figure 1; Ivanova & Leydesdorff, 2013). The virtuous and vicious circles among three (or more) dynamics contain the options of both globalization and localization at each moment of time.

In terms of the Triple Helix, localizations are needed, for example, in order to retain wealth in the (local) economy from global knowledge; but globalization of the perspectives is also needed in order to make an economy knowledge-based. In empirical cases, one can expect a trade-off between local organization (variation) versus global self-organization ("beyond control") in the selection environments of the relevant markets and sciences. In general, the three functional (sub-)dynamics in a Triple Helix can be specified at the systems level as interactions among (i) scientific novelty production, (ii) economic wealth generation, and (iii) political control (Leydesdorff & Zawdie, 2010). Carayannis & Campbell (2009 and 2010) proposed further extensions of the TH to a Quadruple or Quintuple Helix (cf. Bunders *et al.*, 1999; Leydesdorff, 2012).

This complex adaptive system is differentiated horizontally in terms of the three (or more) functional codes of the communication (economic, scientific, political), and vertically in terms of observable relations versus latent functions. The observable exchange relations in the network can also be made the subject of network analysis. The relations, however, are differently distributed in terms of the structural components—universities, industries, and governments— and accordingly what the relations mean can be different from each of these three perspectives.

The specification of how the codes operate as selection environments upon one another requires a systems perspective on the distributions of both relations and non-relations in terms of *correlations*. Specific relations can also be functionally equivalent. The correlations carry the latent functions that can operate synergetically to a varying extent. The synergy is an interaction effect among the distributions: do the functions fit? Other variants and development patterns are possible in niches (Bruckner *et al.*, 1996; Kemp *et al.*, 1998; Schot *et al.*, 2007). In summary, this TH model of synergy becomes more abstract than a network model because a systems perspective is added that operates in terms of the functions of the relations.

## 2. Mutual information and mutual redundancy in three dimensions

The relative frequency distributions of the relations can be expected to contain uncertainty that can be specified using information theory (Shannon, 1948; Theil, 1972). Ulanowicz (1986; 1997) proposed to measure synergy as mutual information in bi- and trilateral relations when developing Ascendency Theory. Ascendency is the organizing power of the emerging next-order level in an ecosystem's trophic network: which combinations of food distributions enable a system to flourish? More recently, Ulanowicz (2009) suggested considering the loop in Figure 1 (above) as potentially auto-catalytic. Participation of a third party (or reagent) may enhance relations between two of them to such an extent that the loop is reinforced by the mediation. In other words, the TH can then slip into another type of communication dynamics.

Can this indicator also be used for the specification of synergy in TH configurations (Ulanowicz, *personal communication*, October 6, 2001)? The mutual information in three dimensions (which is specified in the Appendix) follows analytically from Shannon's (1948) information theory (e.g., McGill, 1954; Abramson, 1963, at p. 129; Yeung, 2008, at pp. 59f.), but has posed an anomaly within this theory because the results of the measurement can be positive or negative. In Shannon's theory information is defined as uncertainty and is necessarily generated as positive when the system operates (because of Shannon's intentional coupling of the information measure $H$ to the Second Law of thermodynamics).[1] Negative information, however, would be a reduction of uncertainty as in a niche. Pockets of relatively less uncertainty are also assumed in Prigogine's (1980) theory of self-organizing systems when order emerges.

Why can a system of three reduce uncertainty? (Burt, 1992; Simmel, 1902; Tortoriello & Krackhardt, 2010). In a system of three (or more) distributions, the correlation between two of them can be spurious upon the third (Sun & Negishi, 2010). For example, when two parents answer the questions of a child consistently, uncertainty can be reduced. The answers of one parent inform the child in this case about the expected answers of the other because the two sources of variation are coordinated at a systems level (that is, the marriage as a latent construct). The child will not have direct access to this next-order sharing of expectations. Note that the consequent reduction of uncertainty cannot be attributed to specific agents, but is generated in a configuration of relations. In the case of a divorce, for example, positive uncertainty among the answers may begin to prevail over the synergy measurable as negative uncertainty (that is, redundancy) in the configuration.

Similarly, in university-industry-government relations a strong existing relation between two of the partners may make a difference for the third. The synergy indicator measures both positive and negative contributions to the uncertainty or, in other words, both the necessarily positive interaction information and the potentially negative correlational information. Krippendorff (2009a) showed that this summary measure cannot be considered as Shannon information; Shannon-type interaction information ($I_{ABC \rightarrow AB:AC:BC}$; cf. Krippendorff, 1980) can be defined differently. However, Krippendorff (2009b) added that interactions with next-order loops—such as overlays—"entail positive or negative redundancies, those without loops do not" (at p. 676).

---

[1] The communication of (Shannon-type) information necessarily adds uncertainty because Shannon (1948) defined information as probabilistic entropy: $H = -\sum_i p_i * \log_2 p_i$. Probabilistic entropy is coupled to Gibbs' formula for thermodynamic entropy $S = k_B * H$. In Gibbs' equation, $k_B$ is the Boltzmann constant that provides the dimensionality Joule/Kelvin to $S$, while $H$ is dimensionless and can be measured in bits of information (or any other base for the logarithm) given a probability distribution (containing uncertainty). The Second Law of thermodynamics states that entropy increases with each operation, and Shannon-type information is therefore always positive (because $k_B$ is a constant).

The overlay of TH relations provides additional loops of information processing to the interacting agents that may thus reduce uncertainty. Too much reduction of uncertainty, on the one side, may lead to insufficient stimulus to innovate. Too much uncertainty, on the other, may make new relations risky. The TH indicator of synergy indicates local reduction of uncertainty within a system (for example, a regional system of innovations) with a negative sign, and opening up to globalization with a positive one. This choice of the signs is technical, but makes the indicator consistent with Shannon's information theory (Leydesdorff, 2010). Leydesdorff & Ivanova (in press) have shown that the indicator measures "mutual redundancy," that is, the extent to which the same information is coded from two (or more) different perspectives.[2] The perspectives can also be considered as containing latent codes of communication that interact in the observable relations.

In summary, mutual redundancy in three or more dimensions enables us to indicate the configurations under study in terms of local integration in terms of relations and synergy versus global differentiation and opening. However, the indicator is not more than a formal instrument. One needs a research design for the appreciation of the results. In other words, interpretation of the results requires a discussion of a TH system in relation to its relevant selection environments (Meyer *et al.*, 2013). As we shall see below in empirical cases, the synergy values can then serve as a heuristics by focusing our attention on intuitively unexpected possibilities.

Because all information measures are based on aggregation (using sigmas; see the Appendix), the synergy can also be decomposed in terms of the dimensions such as geographical regions, or sectors such as medium-tech manufacturing and knowledge-intensive services. The indicator provides the option to test assumptions of systemness such as in national or regional innovation "systems." In recent studies, attempts were made to include institutions without first specifying their national boundaries (Choi, Yang, & Park, 2014 forthcoming) and research domains (Khan & Park, 2013). One can address these questions empirically: how much synergy is indicated regionally or nationally? Is there surplus between regions, nations or sectors? Is this different for sectors in regions? The measurement results can be fully decomposed.

For example, when one considers the innovation systems of regions in Northern Italy such as Piedmont (OECD, 2009) and Lombardy, the industry in the one region may fit more synergistically to the knowledge base of the other, but the political distribution of authority is regionalized (Beccatini *et al.*, 2003; Cooke & Leydesdorff, 2006). In such a configuration, the further development of university-industry-government relations in one of the regions might impede the trans-regional innovation system, but a trans-regional mechanism of political authority across northern Italy is failing. One can thus indicate limitations on synergy generation and from this perspective provide policy advice.

---

[2] In the case of three (or an uneven number of) dimensions, mutual redundancy is equal to mutual information, but in the case of an even number of dimensions, the sign has to be changed.

In the case of Norway, for example, Strand & Leydesdorff (2013) found that synergy among the three functions could be retrieved in relation to high- and medium-tech offshore activities on the west coast, whereas the traditional university centers (Oslo and Trondheim) have remained more distanced from TH models. In China, Leydesdorff & Zhou (forthcoming) found synergy at the level of the 31 provinces by an order of magnitude more than at the lower level of 339 regional prefectures—with the exception of the four municipalities among them that are centrally administered as provinces (Beijing, Shanghai, Tianjin, and Chongqing). Thus, the knowledge-based economy of China seems not coordinated at the lowest regional level. Expanding the model of the four municipalities to other regions might enhance the knowledge-based pattern in the economy.

In Russia, Leydesdorff, Perevodchikov, & Uvarov (in preparation) found that the knowledge-intensive services—which are in Russia often related to state apparatuses—were synergetic at the local level, but did not enhance globalization. In western countries, KIS is more "footloose" (Vernon, 1979) and thus a globalizer: when KIS is needed elsewhere, it travels. Similarly, Asian nations such as India and Indonesia show levels of synergy much higher than China and western nations in a study of UIG co-authorships relations in scholarly publications (Ye *et al.*, 2013). What may be synergetic from one (e.g. industrial) perspective may fail to be so from a perspective of regional government or university development.

## 3. Empirical studies

Using 1,432,401 corporate addresses provided on the CD-Rom version of the *Science Citation Index 2000*, Leydesdorff (2003) first explored the use of the indicator in terms of the address information in the bylines of these (725,354) records.[3] The addresses were distinguished as academic, industrial, and governmental using a dedicated routine (Park, Hong, & Leydesdorff, 2005). How are UIG co-authorship relations distributed in relation to bilateral relations or single-sector attributions?

In this design, documents are the units of analysis. The corporate addresses to each individual document are scored on three variables: U, I, and G that can be zero or one. If U and I are both positive, this co-authorship counts as a UI relation, etc. For each set, one thus obtains aggregated values for all (seven) possible combinations (UIG, UI, UG, IG, U, I, and G). This design was later extended with international co-authorship relations as a fourth category in a Quadruple Helix model (Kwon *et al.* 2012; Leydesdorff & Sun, 2009). One thus obtains three other possible synergies between University-Government-Foreign (UGF), UIF, and IGF, and also the possibility to study the development of the quadruple helix of UIGF relations. Both the Korean and Japanese datasets allowed studying these indicators longitudinally (Figures 2 and 3, respectively). A similar design was also used for studies at the internet. The internet also contains documents and one can count occurrences and co-occurrences of words like "university," "industry," or "government" (Khan & Park, 2011; Leydesdorff & Curran, 2000; Park *et al.*, 2005; Skoric, 2013).

---

[3] These addresses point to 725,354 records contained in this database on a total of 778,446 items. Only 3.7 % of these records contain no address information. The total number of authors in this database is 3,060,436. Thus, on average each record relates to four authors, but at two addresses.

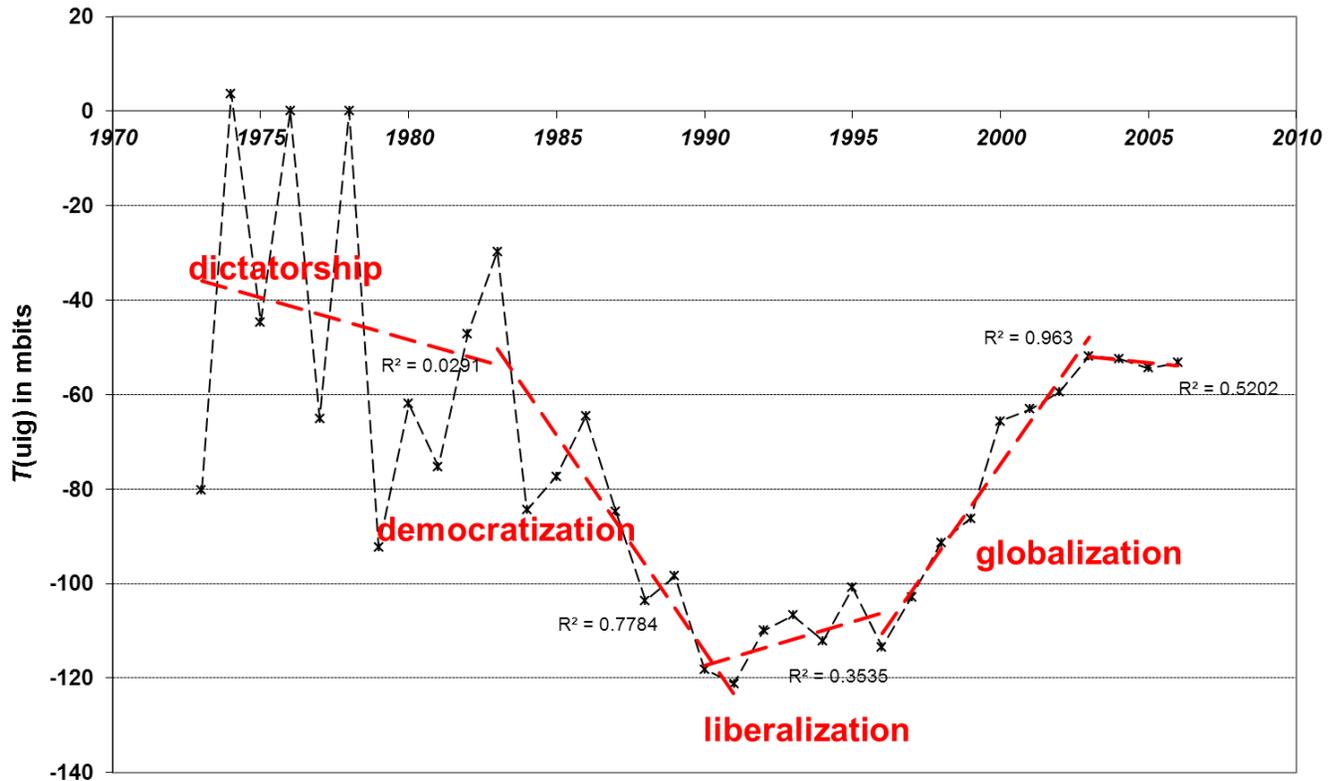

**Figure 2**: Synergy in TH co-authorship relations in Korea 1973-2005. Source: Park & Leydesdorff, 2010: 645; trend lines added.)

Figure 2 shows the result for the national TH system of publications with Korean addresses in the *Science Citation Index* (SCI). Several trend breaches are indicated. Under the dictatorship (until 1987), the increases in synergy in TH relations were initially slow. Democratization and civil liberty stimulated TH development in the period of transition. The 1990s witnessed a reversal of this trend; this reversal was reinforced by the increasing globalization (after, for example, the opening of China). Globalization can be expected to make local interactions among and between sectors less important than sector-specific ones because the latter can be more specialized. During the most recent years, the Korean publication system seems to stabilize again into a balance between local and global dynamics.

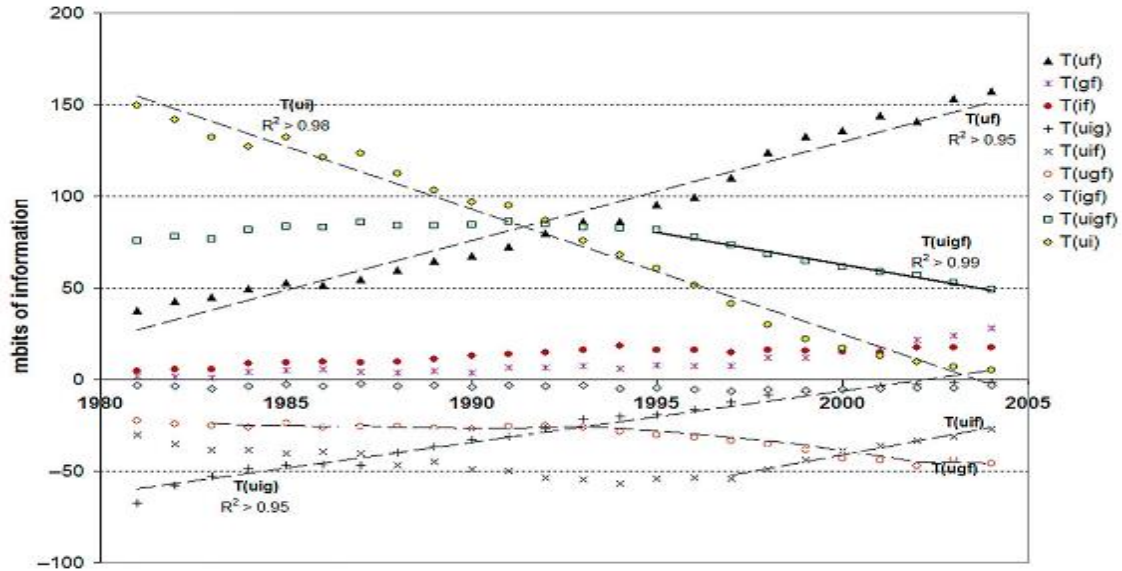

**Figure 3:** The mutual information in two, three and four dimensions among Japanese articles with a university, industrial and governmental address, and international co-authorships. Source: Leydesdorff & Sun (2009, at p. 783).

Using a similar design—with documents in the *Science Citation Index* as units of analysis—Leydesdorff & Sun (2009) found that the TH synergy in the Japanese set ($T_{UIG}$) was –67.4 mbits in 1981, but declined to –1.2 mbits in 2004 (Figure 3). The decomposition shows that University-Industry ($T_{UI}$) relations declined during the entire period (1981-2004), but international co-authorship relations ($T_{UF}$) steadily increased for university-based authors (Wagner, 2008). In other words, international collaboration is a long-term driver for these authors of scholarly articles more than integration with industry at the national level.

The Triple Helix of university, governmental, and non-Japanese authors ($T_{UGF}$) gained in synergy since the first part of the 1990s and this is reflected as a trend breach in the quadruple helix ($T_{UIGF}$). In other words, international co-authorship has played a catalytic role in integrating national TH relations in Japan since the mid-1990s, whereas the national system of co-authorship relations tends to erode as a cultural trend (since the early 1980s). Thus, the effects of globalization are integrated in the Japanese case differently from Korea.

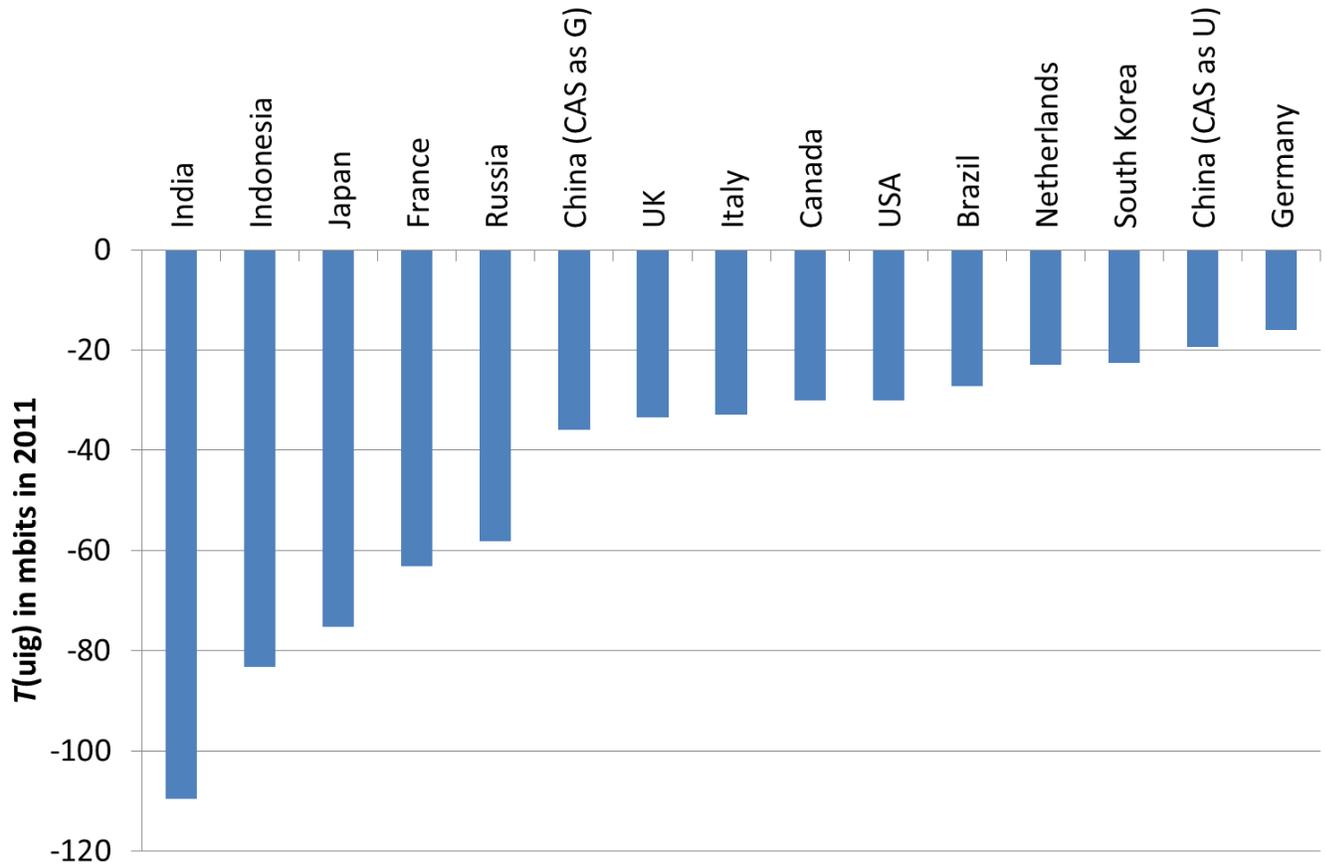

**Figure 4**: $T_{UIG}$ distribution for the various countries in 2011. Source: Ye *et al.*, 2013, p. 2321

Figure 4 shows synergy values in the systems of co-authorship relations in a number of countries, including the G8, using data from *SCI* 2011 (Ye *et al.*, 2013). (The results cannot directly be compared with the results of Figures 2 and 3 because of differences in the data collection). Interestingly, however, Japan and Korea are positioned in Figure 4 at opposite ends. Japan is third after India and Indonesia in terms of national synergy, whereas in Korea the system is globalized at the level of advanced western nations.

One of the objectives of this last study was to consider the position of the Chinese Academy of Science (CAS) which is currently in transition from being a government agency towards being a university-type of institution. When CAS is counted as university, the synergy in the Chinese system is much less (-19.4 mbits) than in the case of counting CAS as governmental (-36.0 mbits). In the former case, China is placed among the most globalized countries (in this domain of publishing) such as Korea and Germany, whereas in the latter case the Chinese system is far more nationally integrated (with a synergy value between Russia and the UK).

In summary, we have witnessed since the mid-1990s an unraveling of the Triple Helix in terms of co-authored publications under the pressure of globalization. Lawton-Smith and Leydesdorff (under submission) note that the Triple Helix model was proposed by Etzkowitz & Leydesdorff (1995) before this change in the system towards globalization became so prevalent. Internationalization was further reinforced by the competitive emphasis on university ranking since the introduction of the Shanghai-rankings in 2004 (Shin *et al.*, 2011). Patenting, for example, is no longer an incentive in leading universities (Leydesdorff & Meyer, 2011; 2013; cf. Etzkowitz, 2013). Institutional incentives have been increasingly focused on international publications and co-authorship relations during the last decade.

**Firms as units of analysis**

Van der Panne & Dolfsma (2003) used micro-data of Dutch firms which were operationalized in terms of the TH dimensions (Storper, 1997): (i) postal codes which indicate the region, etc., in the geographic dimension; (2) technological classifications (NACE codes of the OECD);[4] and (3) firm sizes indicating the economic dynamics (e.g., small- and medium sized enterprises versus large corporations). Mutual information (co-variation) between each two distributions can be used as a relational measure so that synergy in three dimensions can be estimated. Synergy at the national level can then be disaggregated in terms of geographical levels (regions, provinces) or in the other two dimensions (sectors and size classes).

Leydesdorff, Van der Panne, and Dolfsma (2006) published the results of this analysis in a special issue of *Research Policy* dedicated to the Triple Helix. This new application of the indicator in the economic domain led to a series of studies at the national level. Leydesdorff & Fritsch's (2006) decomposition of Germany, for example, was published in the same issue. These two studies led to the conclusion that in the Netherlands additional synergy is indicated at the national level (on top of the sum of the regions and provinces), but in Germany the States (*Länder*; e.g., Bavaria) of the Federal Republic are the most relevant level of integration. Using data of 2004, however, the German *Länder* were still significantly different along the East-West divide, but at the next-lower (NUTS-2)[5] level of regions (*Regierungsbezirke*) the East-West divide was no longer dominant (Figure 5).

---

[4] NACE is an abbreviation of *Nomenclature générale des Activités économiques dans les Communautés Européennes*. The NACE code can be translated into the International Standard Industrial Classificiation (ISIC).
[5] NUTS is an abbreviation for "*Nomenclature des Unités Territoriales Statistiques*" (that is, Nomenclature of Territorial Units for Statistics). The NUTS classification is a hierarchical system for dividing up the economic territory of the EU.

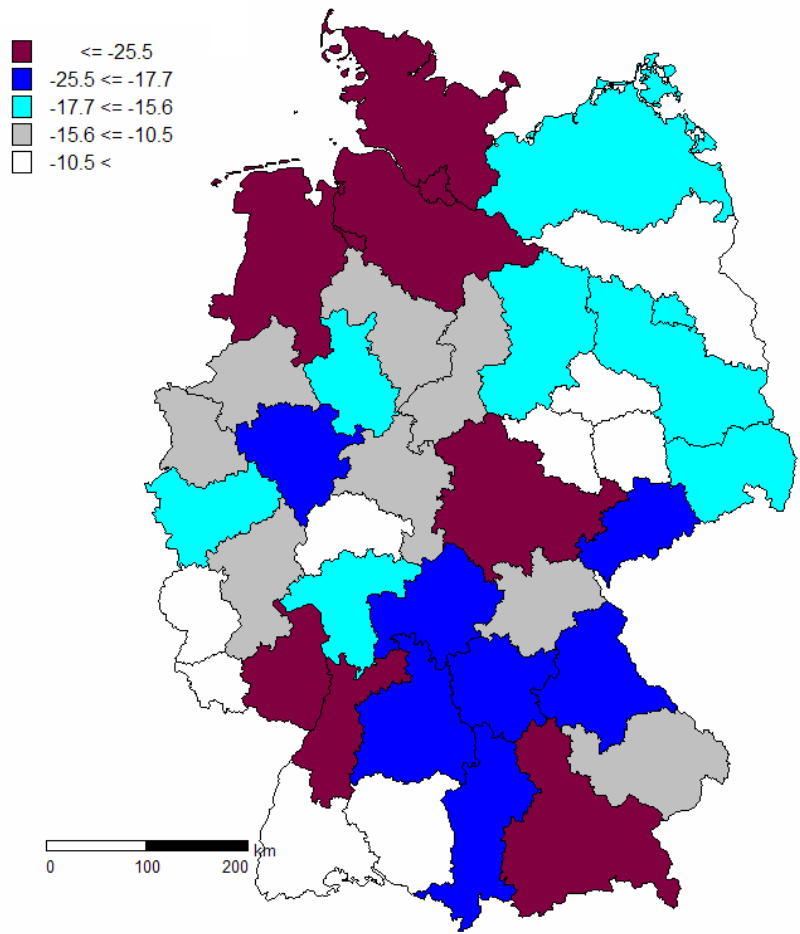

**Figure** 5: The mutual information in three dimensions ($T_{GTO}$) at the NUTS-2 level in Germany in 2004. Source: Leydesdorff & Fritsch, 2006, p. 1544.

Figure 5 shows the synergy at the NUTS-2 level of regions ("*Regierungsbezirke*") in Germany. In terms of sectors, both studies indicated that medium-tech manufacturing is a crucial factor for generating synergy because of its relative embeddedness (Cohen & Levinthal, 1990). High-tech manufacturing and knowledge-intensive services (KIS) can be less embedded, and globalized or footloose (Vernon, 1979). KIS tends to uncouple from local economies because access to an airport or train station is crucial. This uncoupling from the geographical locus is counter-acted when KIS is high-tech because of the need to maintain a laboratory or other specific installation (e.g., computer centers).

The synergy at the regional level was found in Germany around the major metropolitan centers in the western part of the country such as Munich, Hamburg, and Frankfurt; but not in Berlin (cf. Lengyel, Sebestyén, & Leydesdorff, 2013). Thuringia, previously in the German Democratic

Republic, however, is also high on synergy. This is a puzzling result which we only understood after further analysis of the Hungarian innovation system in similar terms. In this follow-up study of Hungary, but using a less complete dataset, Lengyel & Leydesdorff (2011) concluded that Hungary no longer provides surplus synergy at the national level to three regional subsystems that are differently integrated.

The most important subsystem in Hungary was found in the metropolitan area of Budapest and its environment. The western parts of the country were no longer integrated nationally, but had successfully "accessed" to relevant environments in Austria, Germany, and the remainder of the EU, whereas some eastern provinces had remained synergetic under the regime of the old (state-controlled) system. Although these conclusions were tentative, they could be supported by another reading of existing statistics. This disintegration of the national system followed during the 1990s after the demise of the Soviet Union, when the country first went into transition and then became an accession country to the EU. Given the dynamics of European unification, however, it was too late to construct a national system of innovations when Hungary became a free nation.

In a similar vein, Strand & Leydesdorff (2013) concluded on the basis of the full set of Norwegian firm data that the knowledge-base of this national system was no longer primarily integrated in relation to the national universities, but driven by foreign investments in offshore industries in the western parts of the country. Internationalization and globalization in the latter two studies seemed thus core dimensions for understanding how the knowledge-based economy operates. These conclusions, however, made us turn to the Swedish innovation system as a baseline for comparison and validation (Leydesdorff & Strand, 2013).

More than any other, the Swedish innovation system has been studied in detail, is documented statistically with great precision, and could provide us with a benchmark to test our Triple Helix methodology. Using the full set of data about 1,187,421 firms made available by Statistics Sweden (November 2011), the data was analyzed with a design similar to the previous studies. The Swedish case, however, allows us to specify an expectation.

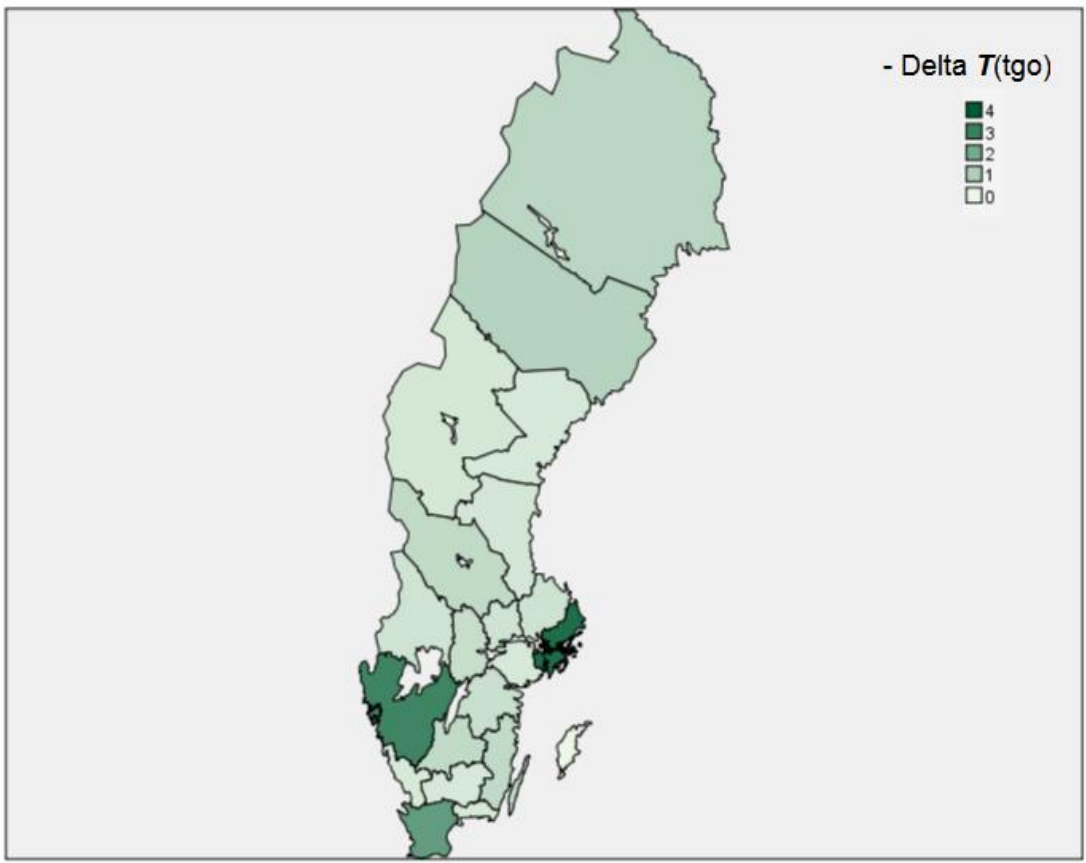

**Figure 6**: Contributions to the reduction of uncertainty at the level of 21 Swedish counties. (Source: Leydesdorff & Strand, 2013, p. 1897).

In accordance with the expectation, three counties dominate the picture of the synergy within this nation (Figure 6) when decomposed in terms of regions: Stockholm (–3.49 mbits), Västre Gotalands län including Gothenberg (–2.91 mbits), and Skåne including Malmö and Lund (–2.31 mbits). Together, these three regions account for (8.71 / 17.95 =) 48.5% of the summed synergies of regions at the geographical scale of NUTS3. The between-region synergy at the national level is –4.61 mbits or 25.7%, so that the other 18 counties contribute (100 – 48.5 – 25.7 =) 25.8% to the national synergy.

All relevant distributions (of bi- and trilateral relations) are more skewed in Stockholm than in Gothenburg and Malmö. Most importantly, the synergy function $\Delta T_{GTO}$ is far more negative for Stockholm. In summary, the different functions are more concentrated and the distributions operate more synergistically in the Stockholm region than around Gothenburg and Malmö/Lund. The other regions follow these three regions at a considerable distance.

The final study which we wish to mention in this context is that of China. In this case we used data downloaded (Dec. 2012) from the *Orbis™* database of Bureau van Dijk (available at https://orbis.bvdinfo.com ). This data at the firm level is collected for commercial purposes; the official statistics of China are not available to the public. The results showed that most of the synergy in the Chinese innovation system is generated at the level of the 31 provinces: only 18.0%—that is, less than Sweden—was generated at the next level of the nation between the provinces.

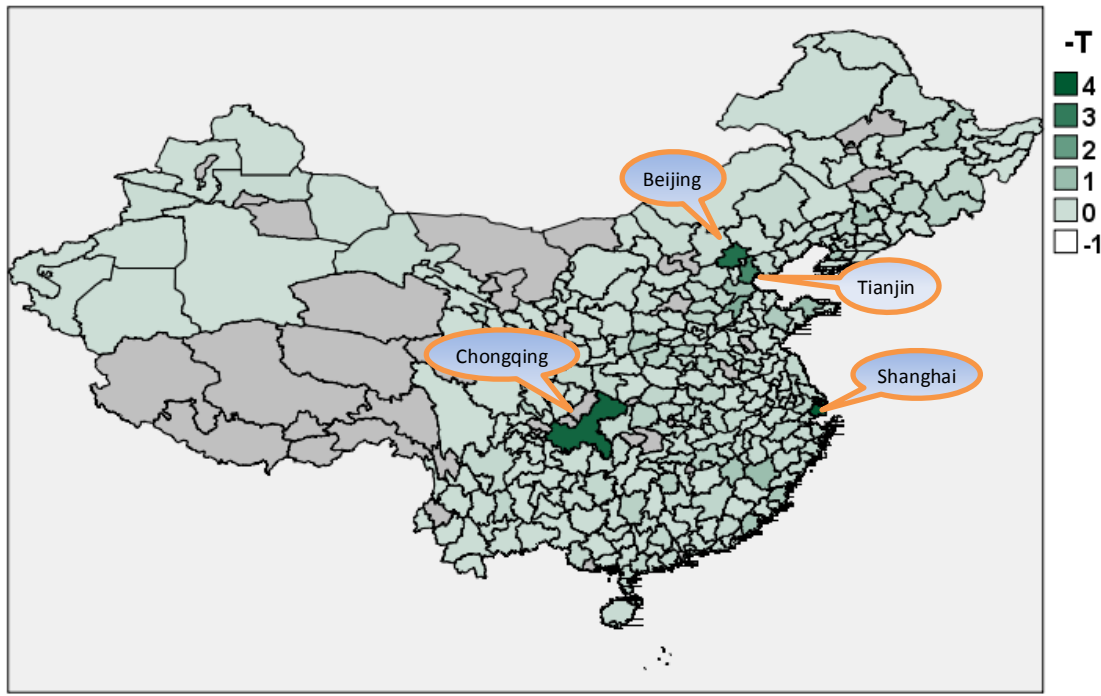

**Figure 7**: The distribution of 339 second-level administrative units in the PRC compared in terms of their contribution to the synergy among technology, geography, and organization. Source: Leydesdorff & Zhou (in press).

Figure 7 shows an interesting finding of this study. At the next-lower level of the 339 prefectures, the four municipalities which are governed centrally as provinces stand out with together 33.0% of the total synergy at this level. This result came unexpectedly from our analysis, but was immediately recognizable to our Chinese colleagues. The success of these four municipalities in constructing a knowledge-based economy regionally can perhaps be used as a role model for other regions.

**Conclusions**

The operationalization of the TH as a measure of synergy in university-industry-government relations was pursued during the last decade using the mutual information in three (or more) dimensions as an indicator. This indicator was also automated as a routine available at http://www.leydesdorff.net/software/th4 that provides values for mutual information in two, three, and four dimensions of the data. Important advancements were made in understanding this indicator as a measure of redundancy (Krippendorff, 2009b; Leydesdorff & Ivanova, in press). Furthermore, the indicator has recently been applied to some complex social contexts: music festivals (Khan, Cho & Park, 2011), socio-ideological divisions (Kim & Park, 2011), social media-mediated innovation (Choi, Park & Park, 2011), and presidential election campaigns (Park, 2013).

When the results of using publications or firms as units of analysis, respectively, are compared, the different dynamics of globalization and localization become manifest. In the system of scientific publishing the TH tends to unravel at the national level because of the increased priority of international publishing in academia. The competition is organized in terms of academic quality, and the nation has become less important as a frame of reference from this perspective. The dimension local-global seems the best candidate for a further extension of the model with a fourth helix, but one should take care that the sign has to be changed in order to calculate the mutual redundancy consistently in a Quadruple-Helix model (see the Appendix; Ivanova & Leydesdorff, in press; Leydesdorff & Ivanova, in press).

Whereas globalization prevails in the domain of scientific publishing, in the economy localization in terms of retaining wealth from knowledge can be expected to generate an opposite dynamics. Metropolitan areas can be expected to provide most synergy because they combine in a micro-cosmos the advantages of localization and globalization. Leydesdorff & Deakin (2011) used in this context the concept of "meta-stabilization": as the large city can be expected to remain in transition between the local and global dimensions, it can reflect the global dynamics of a knowledge-based economy and become itself a localized center of attraction for preferential attachments.

Such cities do not have to be capital cities (e.g., London, Seoul), but can also be "smart cities" such as Montreal, Munich, and perhaps Edinburgh as centers of concentration and mixing of functions (Daekin, 2014). Furthermore, as in the Chinese case, a municipality or province with its own jurisdiction corresponding to a metropolitan area can similarly be a relevant unit of analysis (Shapiro, Park, & So, 2010; Shapiro & Park, 2012). In these contexts, the local uncertainty may be reduced because sources of variation from the global level can be imported and thus strengthen the local economy.

**Appendix**

Mutual information in three (or more) dimensions—the Triple-Helix indicator to be used here—is a signed information measure (Yeung, 2008), and therefore not a Shannon-information (Krippendorff, 2009a and b). However, this measure is derived in the context of information theory and follows from the Shannon formulas (e.g., Abramson, 1963; Ashby, 1964; McGill, 1954). If the 2-base is used for the logarithm all values are in bits of information.

According to Shannon (1948) the uncertainty in the relative frequency distribution of a variable $x$ ($\sum_x p_x$) can be defined as $H_X = -\sum_x p_x \log_2 p_x$. Shannon denotes this as probabilistic entropy, which is expressed in bits of information if the number two is used as the base for the logarithm. (When multiplied by the Boltzman constant $k_B$, one obtains thermodynamic entropy and the corresponding dimensionality in Joule/Kelvin. Unlike thermodynamic entropy, probabilistic entropy is dimensionless and therefore yet to be provided with meaning when a system of reference is specified.)

Likewise, uncertainty in a two-dimensional probability distribution can be defined as $H_{XY} = -\sum_x \sum_y p_{xy} \log_2 p_{xy}$. In the case of interaction between the two dimensions, the uncertainty is reduced with the mutual information or transmission $T_{XY}$, as follows: $H_{XY} = (H_X + H_Y) - T_{XY}$. If the distributions are completely independent $H_{XY} = H_X + H_Y$, and consequently $T_{XY} = 0$.

In the case of three potentially interacting dimensions ($x$, $y$, and $z$), mutual information can be derived (e.g., Abramson, 1963: 131 ff.) as:

$$T_{XYZ} = H_X + H_Y + H_Z - H_{XY} - H_{XZ} - H_{YZ} + H_{XYZ} \tag{1}$$

The interpretation is as follows: association information can be categorized broadly into correlation information and interaction information. The correlation information among the attributes of a data set can be interpreted as the total amount of information shared between the attributes. The interaction information can be interpreted as multivariate dependencies among the attributes. A spurious correlation in a third attribute, for example, can reduce the uncertainty between the other two.

Compared with correlation, mutual information can be considered as a more parsimonious measure for the association. The multivariate extension to mutual information was first introduced by McGill (1954) as a generalization of Shannon's mutual information. The measure is similar to the analysis of variance, but uncertainty analysis remains more abstract and does not require assumptions about the metric properties of the variables (Garner & McGill, 1956). Han (1980) further developed the concept; positive and negative interactions were also discussed by Tsujishita (1995), Jakulin (2005), and Yeung (2008: 59 ff.).

Since all terms in Eq. 1 are composed by addition (using $\Sigma$; see Theil, 1972: 20f.), mutual information in three dimensions can be decomposed into $G$ groups as follows:

$$T = T_0 + \sum_G \frac{n_G}{N} T_G$$

(2)

When one decomposes, for example, a country in terms of its regions (or provinces), $T_0$ is between-region uncertainty or a measure of the dividedness among the regions; $T_G$ is the uncertainty at the geographical scale $G$; $n_G$ the number of firms at this geographical scale $G$; and $N$ the total number of firms in the dataset. The values for $T$ and $T_G$ can be calculated from the distributions (using Eq. 1), when the values of $N$ and $n_G$ are known. The normalized values of the contributions of regions to the national synergy ($\Delta T = \frac{n_G}{N} * T_G$) and the between-group synergy ($T_0$) can then be derived. $T_0$ is equal to the difference between the $T$-value for the whole set minus the sum of the subsets.

Note that $T_0$ can have positive or negative signs, and can also be expressed as a percentage contribution to the total synergy for a system of reference (e.g., at the national level). A negative value of $T_0$ indicates that the uncertainty at the next geographically aggregated level is reduced more than the sum of the parts, whereas a positive value indicates that the next level of integration does not add synergy to the system. Thus, one can test, for example, whether the national level adds to the systems integration more than the sum of the regional units. The relative contributions at each level can be specified after proper normalization for the number of firms.

The extension to more than three dimensions is straightforward, but one has to take care of sign changes between odd and even numbers of dimensions of mutual redudancy in order to remain consistent with Shannon's (1948) mathematical theory of communication (Leydesdorff & Ivanova, in press). The online calculator at http://www.leydesdorff.net/software/th4 provides results in two, three, and four dimensions of the mutual information so that one can also test four-dimensional models (Bunders *et al.*, 1999; Carayannis & Campbell, 2009; Leydesdorff & Sun, 2009).